# CARBON AND BORON NITRIDE NANOSTRUCTURES FOR HYDROGEN STORAGE APPLICATIONS THROUGH A THEORETICAL PERSPECTIVE

Y. T. Singh[a,b], B. Chettri[a,b], A. Banik[c], K. O. Obodo[d] and D. P. Rai[a],*

[a]*Physical Science Research Center (PSRC), Department of Physics, Pachhunga University College, Mizoram University, Aizawl 796001, India*

[b]*Department of Physics, North-Eastern Hill University, Shillong 793022, India*

[c]*Department of Electrical Engineering, National Institute of Technical Teachers' Training & Research (NITTTR), Kolkata, India*

[d]*HySA Infrastructure Centre of Competence, Faculty of Engineering, North-WestUniversity (NWU), P. Bag X6001, Potchefstroom, 2520, South Africa*

**Abstract**: The recent progress in the field of hydrogen storage in carbon and boron nitride nanostructures has been summarized. Carbon and boron nitride nanostructures are considered advantageous in this prospect due to their lightweight and high surface area. Demerits of pristine structures to hold hydrogen molecules for mobile applications have been highlighted by many researchers. In such cases, weak van der Waals interaction comes into account, hence, the hydrogen molecules are weakly bonded with the host materials and hence weak adsorption energy and low hydrogen molecules uptake. So, to tune the adsorption energy as well as overall kinetics, methods such as doping, light alkali-alkaline earth metals decoration, vacancy, functionalization, pressure variation, application of external electric field, and biaxial strain has been adopted by many researchers. Physisorption with atoms decoration is promising for hydrogen storage application. Under this condition, the host materials have high storage capacity with considerable average adsorption energy, feasible adsorption/desorption kinetics.

*****Address correspondence to Dr. D. P. Rai:** Physical Science Research Center, Department of Physics, Pachhunga University College, Mizoram University; E-mail: dibya@pucollege.edu.in

# INTRODUCTION

Population increase, as well as rapid population surge in different parts of the world accompanied with the need for a sustainable and better quality of life ave, resulted in a significant increase in energy demands [1]. Presently, fossil fuels are the primary and dominant source of energy due to their established infrastructure, ease of delivery and cost competitiveness compared to other sources of energy. Fossil fuels are known to cause serious harm to the environment as a result of the emission of harmful pollutants, which is accompanied by their use in different industries. The need for renewable energy sources to mitigate these environmental challenges as well as the securing energy future due to the limited availability of fossil fuels is great. The transition to a clean renewable energy source is important due to the various drawbacks of fossil fuels.

Hydrogen is abundant in nature but not in a free state. It can be utilized as an energy carrier because it does not have any harmful by-products during combustion (by-product is water) and has high energy density compared to other elements [2], [3]. Even though hydrogen has many advantages over other energy sources, the problem lies in its storage. As per the United States Department of Energy, the benchmark hydrogen uptake capacity should be above 6.5wt%. Different means of storing hydrogen is been explored such as compressed gas, liquid organic hydrogen carriers, inorganic systems, etc.[4]–[6] The average adsorption energy benchmark is set to 0.2-0.8 eV per $H_2$ molecule at ambient conditions. However, considering the recent progress made in hydrogen storage materials, they lack at achieving all the benchmark criteria. Also, the current experimental storage methods are not cost-effective and also have safety concerns.

The possibility of hydrogen storage on nanomaterials like silicene, TMDs, carbon nitrides, silicon carbides, boron carbides, metal hydrides, magnesium-based hydrides, metal nitrides/amides/imides, polymers, clathrate hydrates, zeolites, metal-organic frameworks (MOFs) has been studied. Most of these materials have storage capacity in the desired range, but the hydrogen molecules are mostly chemically absorbed with the host materials. In such cases, the materials are not considered fit for onboard applications. At present, researchers are also focused on MOFs, which under doping show a favourable response with better adsorption/desorption kinetics towards the hydrogen storage [2]-[13]. Different one-dimensional nanostructures due to their high porosity are investigated for hydrogen storage applications. Kumar et al. theoretically investigated magnesium functionalized on the boron clusters for hydrogen energy storage application. Boron cluster with 6 boron and 2 magnesium decoration were found to be promising with a gravimetric density of 8.10 wt% [19]. Magnesium oxide(Mg12O12) nanotube showed a higher possibility of surface adsorption of 12 - 24 hydrogen atoms. The adsorption of hydrogen on Mg and O sites decreased the work function of the nanotube [20]. Silicon carbide nanotube when functionalized with transition metals and silicon carbide nanocages are reported to be suitable for hydrogen molecule storage [21], [22].

Carbon nitride nanostructures where the bonding between carbon and nitrogen atoms is formed by $sp^2$ hybridization and are similar to graphene have also garnered interest from researchers. The hydrogen adsorption possibility on such nanostructures has been theoretically analyzed due to its lightweight and high surface area. Wang et al. highlighted the possibility of lithium and calcium co-doped g-CN nanostructures for hydrogen storage application. They reported that the hydrogen molecules are adsorbed on the host material with an adsorption energy of 0.26 eV/H2 and a gravimetric density of 9.17 wt%. The adsorption mechanism is through polarization of the H2 molecules by the transfer of charge from lithium and calcium atoms [23]. $B_2S$ monolayers which also share graphene-like properties show a promising response towards hydrogen storage. Liu et al. reported the gravimetric density of 9.7 wt% on lithium decorated B2S nanosheet with a binding energy of 0.14 eV/H2 using dispersion corrected density functional theory [24]. GeC monolayer, which shows semi-conducting characteristics, is another interesting 2D nanomaterial that is highlighted by Arellano et al. for its hydrogen adsorption and storage ability. They decorated the GeC nanosheet with alkali and alkaline earth metals. Weakly adsorption of alkaline earth metals on the GeC nanosheets discarded its possibility towards hydrogen adsorption whereas, alkali metals are chemically adsorbed on the GeC nanosheet and tunes the hydrogen adsorption ability. Potassium adsorbed GeC nanosheet shows better kinetic towards hydrogen adsorption [25]. Light alkali metals decorated borophene with a gravimetric density of 13.96 wt% was reported recently by Wang et al. [26]. Calcium decorated MoS2 under the application of external field was reported to store the 6 hydrogen molecules by Kubas-interactions with adsorption energy ~ 0.14 eV per H2. Hydrogen molecules are found to be strongly bonded on such bulky systems but are unable to meet the gravimetric density criteria set by USDOE [27].

Carbon nanotubes and Graphene were first experimentally synthesized by Ijima et al. Novoselov et al. back in 1991 and 2004 respectively [28], [29]. Since then, graphene, carbon nanotubes(single and multi-walled) has been highly explored by the scientific community for their applications in device fabrications, sensors, drug delivery owing to their unique chemical, electronics, thermal, and mechanical properties. Bolotin et al. highlighted the high electron transport and mobility in suspended graphene. Later, the scope of graphene for electronics and photonics device application owing to its high charge mobility and electron transport was reported by Avouris et al. [30], [31]. Functionalized graphene and CNTs are proving to be useful in biomedical applications. Their invitro and invivo toxicity upon drug delivery analysis shows a higher possibility of its utilization in many biomedical applications [32]–[37]. Owing to its superior thermal property, graphene has been realized for application in thermal interface materials. The thermal conductivity of the graphene incorporated thermal interface materials was highly improved by a factor of ~20. Such kind of approach was found to be promising to build thermoelectric devices in the future [38]–[40].

Experimental growth of boron nitride nanotubes using arc-discharge was first made possible in the year 1995 by Chopra et al [41]. The tight binding theoretical model predicted the possibility of BN nanotubes in the year 1993 by Rubio et al. and a year later by Blase et al. using ab initio pseudopotential method also predicted the single and multi-walled BNNTs [42], [43]. Boron nitride nanotubes are also considered to be an interesting prospect due to their high young modulus ~ 1.2 TPa, high thermal stability, and chemical inertness. As the defects like vacancies, Stone-wales are common during the growth of nanostructures, defect engineering has been widely followed to tune the properties of the nanomaterial for device applications [44]–[46]. Boron nitride nanosheets are also known as 'white graphene due to the similarity in structural properties was experimentally realized by Novoselov et al. in the year 2005 [47]. As the boron nitride nanosheets are chemically inert, the possibilities of anti-cancer drug delivery using density functional theory and molecular dynamics have been investigated by Vatanparast et al. The adsorption of the anti-cancer drug and their delivery to the target cell is proved to be feasible [48]–[50].

The BNNTs capability for drug delivery application for breast chancer therapy was recently highlighted by Avval et al [51]. Boron nitride nanotubes due to their high porosity and unique partial ionic B-N bonding have been investigated for gas sensing applications. The feasibility of the transition metal doped BNNTs as nitrogen monoxide, carbon monoxide sensors was investigated. Also, ammonia adsorption under the application of an external electric field was reported on the BNNTs [52]–[55]. SW and DW-BNNTs have shown a promising response for water purification. They can capture a higher concentration of methylene blue particles present in the water [56]. Boron nitride nanotube incorporation into thermoplastic materials using solution blending was reported to tune the thermal conductivity. Thus can be utilized as a filler in polymers [57]. Similar to graphene, the possibility of BNNTs was also realized for thermal interface materials [57], [58]. Spin-splitting effects were observed on the open BNNTs. Similarly, the local magnetic moment was introduced on BNNTs and BNNSs upon functionalization, doping by transition metals, or by creating a vacancy at B/N sites such that the BNNTs can be utilized for spintronics devices [59]–[65].

Lightweight solid-state materials with i) high surface area, ii) greater hydrogen molecule adsorption capabilities, provide a better opportunity for hydrogen energy-based onboard application. In this regard, carbon and boron nitride nanostructures are considered efficient and promising for the storage of hydrogen. We have highlighted the recent progress and interesting findings made on the carbon and boron nitride-based nanostructures for hydrogen storage applications by different researchers, research communities through theoretical (DFT) insights.

**APPLICABLE METHODS**

Density functional theory(DFT)[66] implemented computational software such as VASP, Quantumwise ATK, CASTEP, GAMESS, $Dmol^3$, etc. were employed by the researchers to study the hydrogen adsorption properties of the boron nitride and carbon nanostructures [67]–[71]. In DFT, the Kohn-Sham equation is solved considering many-body electron-electron, electron-ion effects. Some of the common approximation methods used are local density approximation(LDA), generalized-gradient approximation(GGA) to deal with the energy exchange-correlation potential [72], [73]. However, as LDA and GGA are said to have some drawbacks in calculating interaction energies between host materials surface and hydrogen molecules, researchers also adopted an advanced method of van der Waals(vdW) dispersion correction methods such as DFT-D2, D3 [22-23].

Different parameters like binding energy, adsorption energy, average adsorption energy, and desorption temperature are calculated to frame the $H_2$ storage capabilities of the solid-state nanomaterials.

An equation to calculate the adsorption energy is given below [76]–[78]

$$E_{ad} = E_{tot}(BN/C + H_2) - E_{tot}(BN/C) - nE_{free}(H_2) \qquad (1)$$

Similarly, the average adsorption energy is calculated using the following relation [24–26],

$$E_{ad} = \{E_{tot}(BN/C + H_2) - E_{tot}(BN/C) - nE_{free}(H_2)\}/n \qquad (2)$$

where the BN/C denotes the carbon and boron nitride nanomaterials host material, and $E_{tot}(BN/C+H_2)$ is the total energy of the hydrogen molecules adsorbed system, $E_{free}(H_2)$ is the total energy of a free $H_2$ molecule, $E_{tot}(BN/C)$ is the total energy of the host material (carbon and boron nitride nanostructures) and n denotes the number of adsorbed $H_2$ molecules on the host materials.

The hydrogen uptake capacity of the carbon and boron nitride nanomaterials can be calculated using the following relations [76]–[78],

$$H_2(wt\%) = \frac{nM_{H_2}}{(nM_{H_2} + M_{Host})} *100 \qquad (3)$$

where, $M_{H_2}$, $M_{Host}$, and n are the masses of $H_2$, host material (Boron Nitride and Carbon nanostructure), and the number of $H_2$ molecules, respectively.

To quantitatively analyze the desorption process, the desorption temperature ($T_D$ (K)) is estimated using the van't Hoff's equation [79],

$$T_D = \left(\frac{E_{ads}}{K_B}\right)\left(\frac{\Delta S}{R} - \ln P\right)^{-1} \qquad (4)$$

where, $K_B$ is the Boltzmann Constant ($1.38 \times 10^{-23}$ J K$^{-1}$), $\Delta S$ = 75.44 J K$^{-1}$ Mol$^{-1}$ is the $H_2$ entropy change from gas to the liquid phase at equilibrium pressure P=1atm and R(= 8.31 J K$^{-1}$ Mol$^{-1}$) is the gas constant.

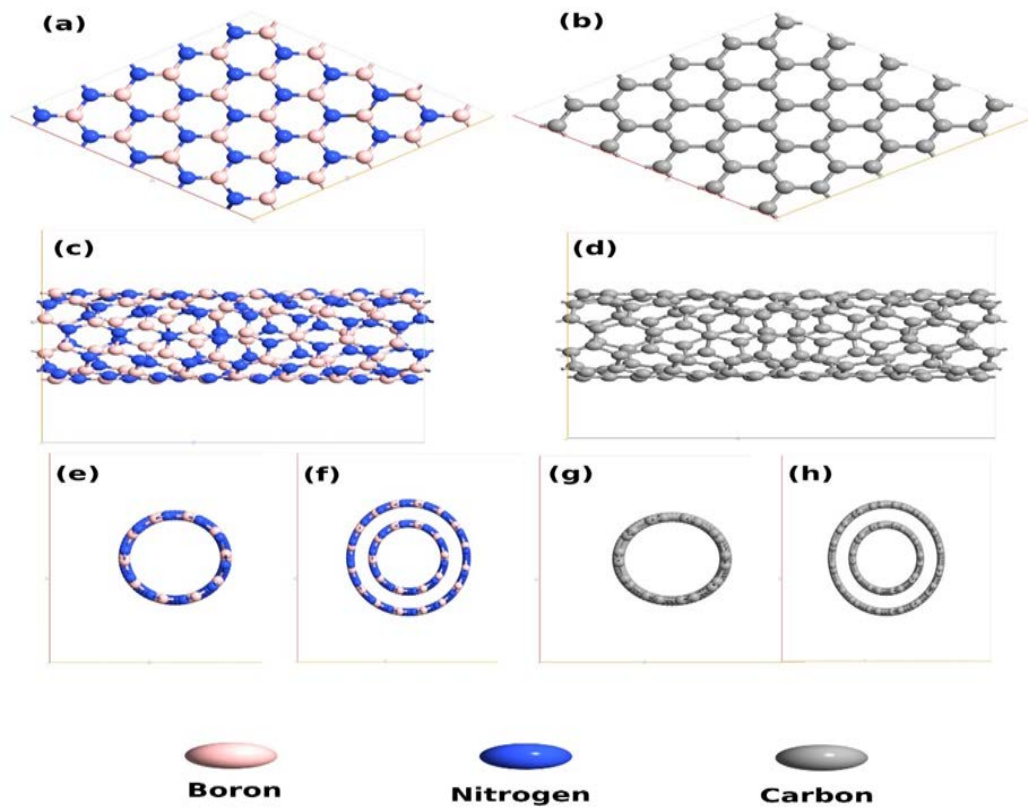

**Figure 1:** Boron nitride and carbon nanostructures. (a), (c), (e), and (f) are Boron nitride nanosheet, side view of single-walled boron nitride nanotube, front-view of single-walled boron nitride nanotube, and double-walled boron nitride nanotube. Similarly, (b), (d), (g), and (h) are graphene, side view of single-walled carbon nanotube, front-view of single-walled carbon nanotube, and double-walled carbon nanotube.

## CARBON NANOSTRUCTURES

Large hydrogen storage capacity in materials like metal hydrides and complex metal hydrides have been reported in many research papers [80]–[82]. But adsorption of hydrogen in such materials mainly happens through the chemisorption mechanism which results in high absorption energy. Such high adsorption energy leads to poor reversibility avoiding the desorption process [82], [83]. Carbon nanomaterials like graphene and carbon nanotubes are other promising materials for solid-state hydrogen storage as they possess a large specific surface area, high polarity, and low mass density [85]. Graphene is a single layer hexagonal lattice of $sp^2$ hybridized carbon atoms. It is semi-metallic and is a zero bandgap semiconductor (see fig: 2(a)). Early research reports wide applications of graphene in the field of nanoscience it has high thermal stability, flexibility, conductivity, and also high storage capacity [86]. Theoretical and experimental investigations report graphene as a good choice for gas adsorption and desorption due to its low binding energy value of approximately 0.2 eV which can be easily reversed [87]–[89].

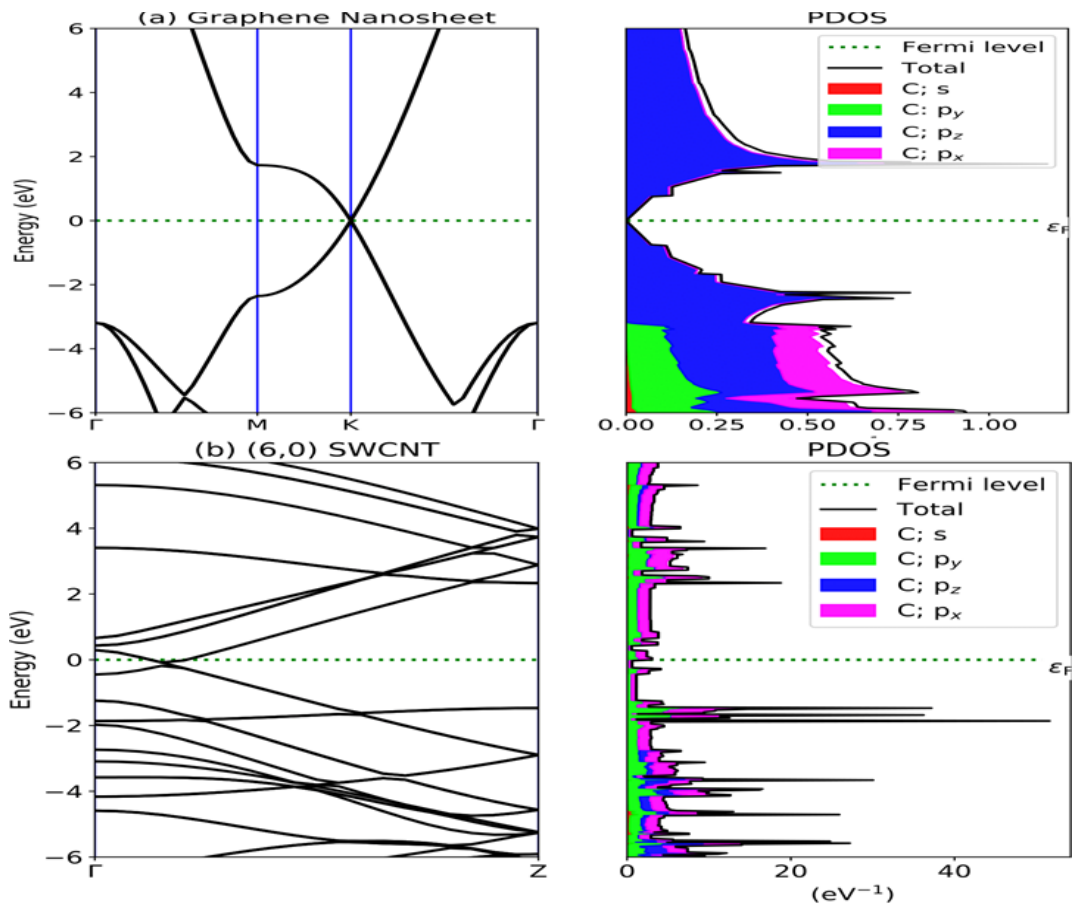

**Figure 2:** (a) Electronic band structure and Projected Density of states for monolayer graphene (b) Electronic band structure and Projected density of states for (6,0) SWCNT.

Carbon nanotubes (CNTs) are another form of carbon-nanomaterials which structure is the same as rolling graphene sheets in some specific direction, but in this case, the carbon atoms are $sp^3$ hybridized. Base on formation, CNTs are of two types: single-walled carbon nanotubes (SWCNTs) and multi-walled carbon nanotubes (MWCNTs). Unlike in the case of monolayer graphene, single-walled carbon nanotubes (SWCNTs) can be either metallic or semiconducting depending on the chirality (n, m) and diameter of the tube [90]. Fig 2(b) shows the band structure and projected density of states (PDOS) of a typical SWCNT.

Investigations by modifying the electronic and mechanical properties have been performed by many researchers and reported its wide applications in making neon devices like nano detectors [91], [92], nano transistors [93], [94], etc. Many researchers also reported CNTs as a promising candidate from gas sensing devices as it shows good selectivity and sensitivity on different gases [95], [96]. Here in this section, we are giving a brief review of the previous work on hydrogen absorption of graphene sheets and CNTs.

**GRAPHENE NANOSHEETS**

Hydrogen adsorption in graphene takes place either by Physioroption or Chemisorption. The physical absorption of hydrogen in graphene shows weak adsorption capacity and also even suggest unsafe by some researchers due to the absorbents' weak stability, which makes it not possible for long time storage [97], [98]. Theoretical investigations reported the maximum gravimetric density (GD) value for graphene due to physical and chemical adsorption as 3.3% and 8.3% respectively [99]. These GD values for graphene change with temperature and pressure. Lai-Peng et al. [100] investigated the hydrogen adsorption behaviour for graphene at the temperatures above critical values taking pressure reference from zero to 100KPa. The authors reported an increase in adsorption capacity with pressure but a decrease when temperature increase. A similar result was also reported by Kim et al. [101]. A theoretical investigation on a bilayer graphene system based on the post-Hartree-Fork/empirical potentials reports enhancement of the storage capacity with an increase in inter-layer spacing [102]. The investigation further reported that at the inter-layer separation of 6 Å to 8 Å the GD value of the system exceeds approximately 30% to 40% of that of the monolayer graphene. Sarah et al. [103] theoretically investigated the influence of curvature on hydrogen storage for mono-layer graphene. The investigation revealed on adsorption of hydrogen more preferably at the local curvature with maximum convex and also reported the high tendency of the hydrogen atoms adsorbed at the minima local curvature to desorb at low temperature.

Storage capacity for pristine graphene reported so far from the theoretical and experimental analysis shows very poor values [100]. Having weak binding energies (0.01ev to 0.06 eV for physical adsorption and 0.67 eV to 0.77 eV for chemical adsorption) is the main reason for such poor storage capacity values. Several investigations have been performed to enhance the storage capacity value. The first principle calculation of the hydrogen adsorption capacity of pyridinic N-doped graphene results in the physisorption binding energy value of 0.64 eV and chemisorption binding energy value 2.03 eV (almost three times of pristine graphene) [104]. Z. M. Ao et al. [97], from theoretical analysis, reported the storage capacity value of 5.13 wt% for aluminium-doped graphene structure at ambient temperature and pressure, which is quite closed to the US DOE value. Graphene with Stone-Wales defect when decorated with Li atom in a field-free condition and the presence of external field exhibits high storage capacity within the average physisorption energy of 0.1-0.6 eV/$H_2$ with a desorption temperature of 289 K [105]. The details of the above investigations are summarized in Table 1.

Table 1: Hydrogen storage capacity, physisorption binding energy, chemisorption binding energy, desorption temperature for different graphene-structure are summarized. N: Nitrogen, Al: Aluminum, $T_D$: Desorption temperature

| Graphene System | Storage capacity (wt%) | Physisorption binding energy (eV/$H_2$) | Chemisorption binding energy (eV/H) | $T_D$ (K) | Ref. |
|---|---|---|---|---|---|
| Pristine-graphene | - | 0.01-0.06 | 0.67-0.77 | - | [100] |
| Pyridinic N-doped graphene | - | 0.64 | 2.03 | - | [104] |
| Al-doped-graphene | 5.13 | 0.260 | - | - | [97] |
| Li-decorated (55-77)-graphene | - | 0.1-0.6 | - | 289 | [105] |

**CARBON NANOTUBES**

Theoretical and experimental analysis on the hydrogen adsorption of CNTs has been performed by many researchers [18], [32], [106], [107]. Early results show inconsistent storage values. Some researchers even reported CNTs as not a good choice for hydrogen storage [108], [109]. Unclear adsorption mechanism (whether Physisoroption or Chemisorption) is the main factor for such inconvenient results. But at present, it is widely accepted that the hydrogen absorption mechanism in CNT is the co-existence of irreversible chemisorption and reversible physisorption [106], [110]. Safa et al. [111], from their investigation through elastic recoil detection analysis, reported the predominance of the physisorption mechanism at cryogenic temperature. If the temperature is in the range from $30^{o}C$ to $100^{o}C$, hydrogen desorption is more common, but at the temperature range from $100^{o}C$ to $300^{o}C$, chemisorption is predominant. Several investigators have also reported the effect on hydrogen storage capacity by internal factors such as specific surface area, tube diameter, tube-curvature, etc., and external factors like methods of measurement, temperature, pressure, etc. [108], [112], [113]. Arellano et al. [25], from the theoretical investigation of the hydrogen storage capacity on SWCNTs of several diameters, reported the increase in molecular hydrogen ($H_2$) binding energy with tube diameter decreases. The same result was also reported by Mpourmpakis et al. [114] and Kentaro et al. [115]. Wang et al. [116], from theoretical analysis of different structures of fully hydrogenated CNTs, reported the inverse square relation of hydrogenation energy and the tube diameter. The author also reported that for the tube of the same diameter, the armchair structure has more $H_2$ binding energy than the zig-zag. The chirality of CNTs affects the storage capacity only in the case of chemical adsorption [110], [114]. But in the physisorption mechanism, the effect for similarity of the tube is very less [106].

A comparative study of isolated SWCNTs and SWCNT bundle on hydrogen adsorption capacity was investigated by Ghosh et al. [117] through Molecular Dynamics Simulation. The authors reported that the adsorption capacity of an isolated SWCNT is significantly higher than that of the SWCNT bundles, at lower temperatures. It is due to the smaller inter-tube spacing distance than the thickness of the adsorbed hydrogen layer, as hydrogen molecules are adsorbed in multilayers at low temperatures. But at high temperatures, SWCNTs bundles show higher hydrogen storage capacity than the isolated SWCNTs. At high temperatures, only a single layer of hydrogen is adsorbed around the nanotubes, and hence adsorption within the interstitial space of the bundle becomes possible. The storage capacity increases with an increase in the interstitial spacing of the tubes. The authors further investigated the storage capacity for square array and triangular array tubes and reported that square array tubes have higher storage capacity than triangular array tubes. The same result was also reported by Muniz et al. [110] et.al. Through the grand canonical Monte Carlo (GCMC) simulation, Muniz et al [118] theoretically investigate the hydrogen storage capacity for a structurally optimized array of SWCNTs. The authors reported that, at a pressure above 1MPa, the square lattice structure has a higher storage capacity than the triangular lattice structure and at the pressure below 1MPa, the triangular lattice structure shows a higher storage capacity value. For the tubes of the same diameter, the storage capacity in SWCNTs is always higher than the MWCNTs. For MWCNTs of the same wall-to-wall spacing, an increase in the number of walls leads to a decrease in storage capacity value. But storage capacity is increased if the wall-to-wall space increases, keeping the number of layers constant [119]. The effects of temperature and pressure on the $H_2$ storage of CNTs have also been reported by many researchers [120]–[123]. At constant pressure, the storage capacity decreases with an increase in temperature, and at a constant temperature, the storage capacity increases with an increase in pressure. Zhao et al [124] experimentally investigated the hydrogen storage capacity setting the reference temperatures at 77 K, 203 K, and 303 K, keeping the pressure at a constant value of 10MPa. The authors reported the decrease in storage capacity as the temperature increases with the maximum storage value of 1.73% at the temperature of 77K. The same result was also reported by Poirier et al. [125]. Considering moderate pressure of 4, 8, 12, 16, and 20 bar at room temperature, Kaskun et al. [126] experimentally studied the effect of pressure on hydrogen storage in Ni-MWCNTs and reported that the storage capacity increase with an increase in pressure which supports the result reported by Darkrim et al. [127].

Theoretical and experimental investigations on the $H_2$ storage of pristine CNTs give small storage capacity values [121], [128]. So far, different researchers have proposed different methods to enhance the storage capacity [129]–[131]. Doping is the most common method for carbon materials. Jung Hyun Cho et al. [132], from theoretical calculation, reported the storage capacity of 2.5 for Si-doped SCNTs which is double that of undoped SWCNTs. Wang et al. [133], base on the first principle study, calculated the hydrogen storage capacity of Al-doped SWCNTs. The authors reported a very high $H_2$ storage capacity value of 28 wt% at room temperature which is significantly higher than the US DOE value. Surya et al [134], from the first principle study, reported the storage capacity of 13.2 wt% for ammonia doped (5,5) SWCNT with small adsorption energy which suits the release of hydrogen molecules at ambient condition. The details of the above investigation are summarized in Table 2.

**Table 2:** Hydrogen storage capacity, average adsorption energy, desorption temperature for different carbon nanotube structures are summarized. SWCNT: single-walled carbon nanotubes

| System | Storage capacity (wt%) | Average adsorption energy(eV/$H_2$) | Desorption temperature (K) | Ref. |
|---|---|---|---|---|
| Pristine-SWCNT | 1.4 | - | - | [132] |
| Si-doped-SWCNT | 2.5 | - | - | [132] |
| Al-(7,7)SWCNT | 28 | 0.131 | 171 | [133] |
| (5,5) SWCNT-5($NH_3$+5 $H_2$) | 8.18 | 0.091 | 148 | [134] |
| (5,5) SWCNT-10 ($NH_3$+5 $H_2$) | 13.2 | 0.092 | 148.5 | [134] |

## BORON NITRIDE NANOSTRUCTURES

Boron nitride(BN) falls in the category of group III-V compounds. It exists in 3 phases: i) hexagonal boron nitride (h-BN), ii) Cubic-boron nitride (c-BN) and iii) Wurtzite boron nitride (w-BN). Hexagonal boron nitride is considered to be the most stable phase. Boron nitride nanotubes (BNNTs), boron nitride nanoribbons(BNNRs), and boron nitride nanosheets (BNNSs) are some common hexagonal BN nanostructures. The boron(B)-nitrogen(N) bond length lies between 1.44-1.50 Å, the distance between B(N)-B(N) is 2.52 Å, and the bond angle between B-N-B(N-B-N) is 120°. Atomic structures of the BNNS and BNNTs are presented in Fig.1.

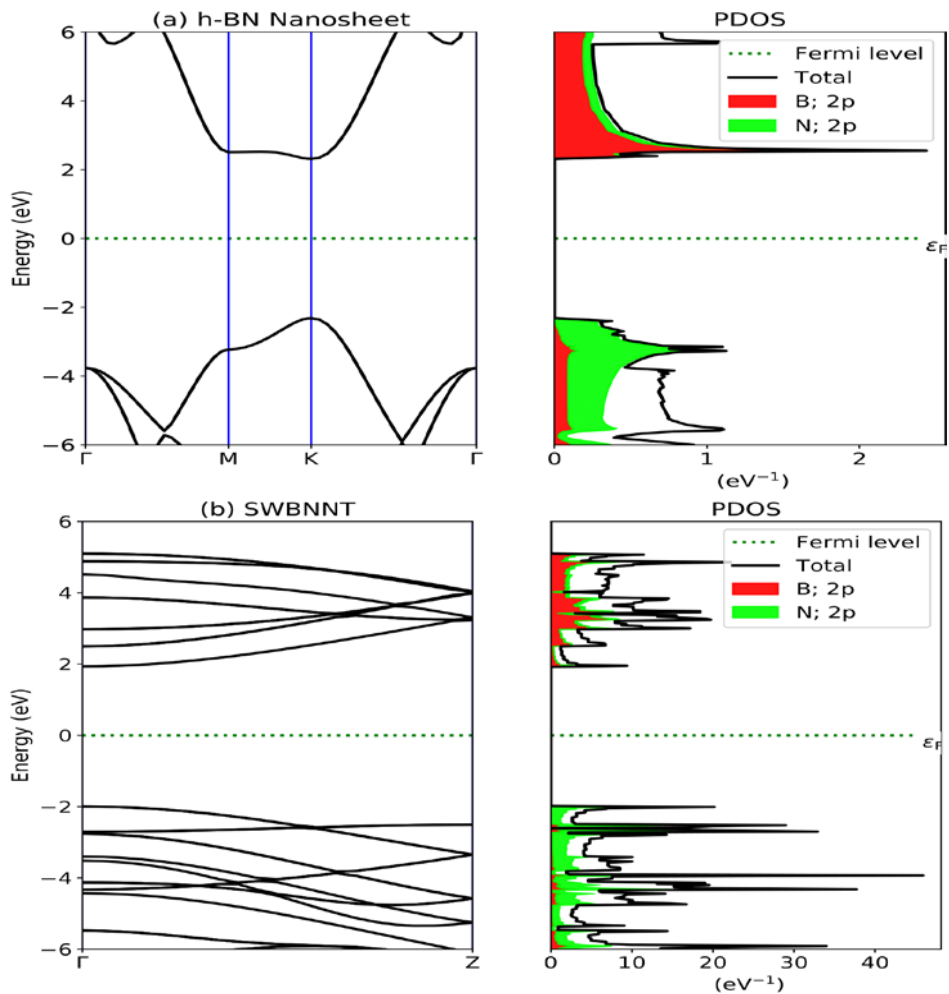

**Figure 3:** Energy band structure and partial density of states for (a) pristine h-BN nanosheet and (b) single-walled boron nitride nanotube. In PDOS red and green color stacked represents boron(B) and nitrogen(N) 2p orbital contribution.

The electronic profile of the boron nitride nanosheet and single-walled boron nitride nanotubes are presented in Fig.2a,b. In literature, the energy band gap of boron-nitride nanostructures lies between 3.6-6.2 eV [135]–[139]. The results are reported from different theoretical and experimental studies. The energy bandgap for pristine boron nitride nanosheet is reported to be 4.67 eV using GGA-PBE as exchange-correlation functional. Whereas by employing hybrid functional(HSEO6), a bandgap tunes to 5.67 eV [137]. In most cases, the valence and conduction bands lie at the K-symmetry point, showing a direct bandgap nature [see Fig. 2a]. The valence band is populated by the N-2p orbitals and the conduction band by B-2p orbitals as seen from the PDOS plot adjacent to the band plot for the boron nitride nanosheet. Similar electronic band profiles are reported for boron nitride nanotubes. A direct bandgap of 3.96 eV along the gamma point is presented in Fig. 3b for a single-walled boron nitride nanotube. In the case of pristine boron nitride nanostructures, the analysis of the electronic profile reveals that the valence and conduction bands are populated by N and B-2p orbitals respectively. Also, a similar trend is followed in the case of multi-walled boron nitride nanotubes with different chiralities [140], [141].

**BORON NITRIDE(h-BN) NANOSHEETS**

Boron nitride shares similarities in structural phases with carbon nanostructures. Like graphene nanosheets, carbon nanotube (CNT) there exists h-BN nanosheets, h-BN nanotubes sharing similar structural properties. These nanomaterials have a high surface area to volume ratio and are lightweight, advantageous in many prospects for hydrogen storage. A common method of hydrogen molecule decoration on the pristine host materials is employed to examine the interactions between them. However, due to poor kinetics between pristine systems and H2 molecules, the pristine systems are unable to attain all the criteria put forward by the USDOE. Hydrogen storage on BNNSs was first experimentally reported by Lei et al. They reported a maximum storage capacity of 5.7wt% under 5 MPa at room temperature. They also checked the desorption kinetics, where ~90% of the H2 molecules released at ambient conditions [142]. Chettri et al. reported the hydrogen storage capacity of BNNSs to be 6.7 wt% under the physisorption process with an average adsorption energy of 0.13 eV/H2. The absence of hybridization between B, N-2p orbital of nanosheet and H-1s orbitals of H2 molecules hints the adsorption process to be through weak van der Waals mechanism [137]. Similarly, anti-Kubas interaction during the hydrogen adsorption process was reported by J. Zhou et al., where they highlighted the hydrogen storage prospect of h-BN nanosheets under the application of external electric field normal to the plane of h-BN nanosheet. It was reported that the H2 molecule's storage could be tuned under applications of the externally applied field due to the polarisation of H2 molecules when the molecules are oriented perpendicular to the plane of h-BN nanosheets [143]. They also reported the possibility of hydrogenation at different B/N and hollow sites as a mode to tune the energy bandgap of the *h*-BN nanosheet in their earlier theoretical work [144].

As most of the physical adsorption process is through a combination of weak van der Waals interaction and minimal charge transfer, the adsorption energy and the desorption temperature criteria are not met most of the time. Hence, some external influences like creating a vacancy, doping, decoration of atoms, applications of strain, external electric field are proved to be promising to enhance the hydrogen adsorption ability. This external influence increases the number of adsorbed hydrogen molecules as well as enhances the binding energy by an increase in charge transfers between the host materials and H2 molecules. Tang et al. studied the effect of external electric field and biaxial strain on the hydrogen storage of Na-decorated boron nitride nanosheets. The external influence of field and biaxial strain increases the number of hydrogens adsorbed by the Na atoms. With a biaxial strain of 15% and external field strength of 5.14 nm-1, Na-decorated BN-nanosheets meet the benchmark criteria for hydrogen storage with an adsorption energy of 0.23 eV/ $H_2$ and hydrogen uptake capacity of 9 wt%. The external factor facilitates the flow of electrons from the hydrogen molecules to the Na-atom. Thus the results highlight that Na-decorated BNNSs are efficient in adsorbing a large number of $H_2$ molecules [145]. A similar approach, where the external electric field is applied to tune the hydrogen adsorption kinetics of the boron nitride bilayers was followed by Chettri et al. They reported the gravimetric density of 6.7 wt% on bilayer boron nitride with adsorption energy in the acceptable range. The external field helped to tune the desorption temperature for such a bilayer boron nitride system with the enhancement in adsorption energy [146].

Hydrogenated boron nitride nanosheet with lithium (Li) functionalization shows a promising response towards $H_2$ storage. On replacing H-atom with Li from the hydrogenated BN-sheet, the Li atom becomes cationic, and the Li-substituted region acts as a binding site for $H_2$ molecules. Thus, Li functionalization improves the hydrogen uptake ability of the hydrogenated BNNSs. The hydrogen molecules are adsorbed through the physisorption process [147]. Boron nitride atomic chains with different length decorated with Li atoms at both ends was seen to be a promising candidate with an $H_2$ storage capacity of 29.2 wt% [148]. Another promising approach is to tune the overall adsorption/desorption kinetics by functionalizing the boron nitride nanosheets by $OLi_2$ and $CLi_3$. Due to the strong polar nature of oxygen and lithium, carbon and lithium, cationic lithium, helps in adsorbing more $H_2$ molecules. As a result, the storage capacity increases to 6.80 wt% with desirable adsorption energy(0.15 eV per $H_2$) [149].

Hybrid structures have also garnered interest to check their hydrogen storage abilities owing to their unique physical and electronic properties. In hybrid structures, the charge distribution is different compared to its constituent pristine structures. As reported by Hussain et al., lithium-decoration on hybrid domains of boron nitride and graphene show promising hydrogen storage capability with a gravimetric density of 8.7 wt%. The hydrogen storage mechanism is driven by the hybridization of Li-p and H-1s orbitals. Transfer of charge takes place from lithium to hydrogen molecules, thus tuning the electrostatic interactions between them. In this way, Li decoration tunes the hydrogen storage ability with an orbital interaction of the hybrid boron nitride and graphene domains [150].

Doping, decoration by foreign atoms has been considered another mode of enhancing and fulfilling the criteria for storing hydrogen molecules on the h-BN nanosheets. Ren et al. reported that Platinum and Palladium doped h-BN nanosheets could store hydrogen up to 4.5 wt% with an average adsorption energy of 1.010 and 0.705 eV per hydrogen molecule. The Kubas interaction and polarisation mechanism for $H_2$ adsorption on BN nanosheet are responsible for the higher adsorption energy [151]. Tokarev et al. presented a detailed study of pristine and oxygen-doped boron nitride for hydrogen storage. They reported that oxygen doping tunes the adsorption capability [152]. Recently, Rad et al. theoretically discussed the possibility of nickel decorated boron nitride nanoclusters as a hydrogen storage material. Decoration of nickel leads to the strong binding of the $H_2$ molecules in their vicinity, thus increasing the hydrogen storage capacity [153]. Kumar et al., using the dispersion corrected DFT reported the enhanced $H_2$ storage ability of the carbon and oxygen co-doped BN-nanosheet [154]. Titanium decorated BN nanosheets have a storage capacity of 6 wt%. The adsorption energy lies in the range 0.3-0.7 eV/$H_2$ [155].

## BORON NITRIDE NANOTUBES

BNNTs are superior to CNTs in many aspects like thermal and mechanical stabilities, chemical inertness, unique bonding nature, which makes it an interesting prospect for energy storage applications. The theoretical progress made to study the credibility of BNNTs to store $H_2$ molecules in recent years is quite outstanding. The hydrogen uptake capacity on (multi-walled and bamboo-like) BNNTs was first reported experimentally by Ma et al. The respective storage capacity of the BNNTs were 1.8 and 2.6 wt% under 10 MPa and at ambient conditions [156]. Later, Jhi et al. performed a theoretical (DFT) study and reported that the modification of the $sp^2$ type bonding can lead to higher hydrogen storage with enhanced binding energy. Also, BNNTs have higher binding energy for $H_2$ molecules compared to CNTs [157]. Boron nitride nanotubes are preferably better hydrogen adsorption materials than carbon nanotubes (CNTs) due to the bonding between B-N atoms. These bonds are covalent and partially ionic. Whereas C-C bonds in the case of carbon nanostructures are covalent. Thus, ionic bonding plays a vital role in the hydrogen molecules' adsorption in BNNTs compared to CNTs. This was validated by the DFT study performed by Mpourmpakis et al. They investigated the $H_2$ storage properties on (5,5) and (9,9) BNNTs and CNTs. Storage capacity and the adsorption energy were reportedly to be higher in BNNTs [158].

Hydrogen molecules adsorption on the single, double and triple-walled BNNTs with different chirality and diameters using grand canonical Monte Carlo (GCMC) Simulation was reported by Ahadi et al. The role of external pressure on the $H_2$ storage performance of all the BNNTs was also reported. Single and double-walled exhibits higher hydrogen molecule adsorption compared to triple-walled BNNTs. The percentage of hydrogen adsorption in the case of multi-walled BNNTs depends on the inner diameter. The larger the diameter, the small is the curvature, and higher is the binding of the hydrogen molecule. Whereas, in single-walled BNNTs, adsorption is high when the nanotube diameter is small. The external pressure in units of MPa tunes the storage capacity of the BNNTs similar to CNTs [159]. As reported by Wu et al., for pristine BNNTs, the hydrogen molecules are adsorbed through surface interaction. When defects like Stone-Wales, substitutional, and vacancy are considered, the hydrogen molecules dissociate and are chemically adsorbed on the BNNTs [160]-[161]. The change in diameter leads to the change in B-N bond-length between and hence a slight variation in the sp2 type bonding, which in turn enhances the reactivities of boron and nitrogen atom, thus tuning the hydrogen adsorption ability on those B and N sites [162]. Han et al. also observed hydrogen molecules' interaction with single-walled BNNTs [163].

**Table 3:** Hydrogen storage capacity, average adsorption energy, desorption temperature for different boron nitride nanostructures are summarized. BNNS: boron nitride nanosheet, BNNTs: boron nitride nanotubes, SWBNNT: single-walled boron nitride nanotubes.

| Boron nitride System | Storage capacity (wt%) | Average adsorption energy(eV/$H_2$) | Desorption temperature (K) | Ref. |
|---|---|---|---|---|
| Expanded hexagonal boron nitride (eh-BN) | 2.46 | - | 243 | [164] |
| Co-doped h-BNNS | 12.94 | 0.18 | 243 | [165] |
| Pt-doped h-BNNS | 4.93 | 0.23 | 271 | [165] |
| Porous BNNS | 7.50 | 0.16 | | |
| Ce-doped BNNTs | 5.60 | 0.22 | 268 | [166] |
| C-doped $(BN)_{12}$ cages | 7.43 | - | - | [167] |
| Ti decorated (8,0) SWBNNT | 3.9-5.7 | - | - | [168] |

The mechanical properties of the BNNTs are reported to be modified under hydrogen molecule adsorption. Under hydrogen molecules adsorption, the buckling strength reduces by an average of 14%, and with the temperature variation of 300-3000K, the buckling strength varies from 13-15% [169]. Single-walled BNNTs are decorated with rhodium, palladium, and nickel to increase the H2 storage capacity. In this case, hybridization between d-orbital of the Rh, Pb, and Ni and H-sigma orbital takes place. Thus, the functionalization of BNNTs leads to an increase in average adsorption energy as well as the hydrogen uptake capacity [170]. Mananghaya et al. investigated BNNTs with vacancy defects and decorated by Transition Metals. Only titanium (Ti) and vanadium(V) functionalization were thermodynamically stable, and a single Ti and V could bind seven $H_2$ with an uptake capacity of 7.2 wt% at ambient conditions [171]. However, chemisorbed hydrogen molecules are strongly bonded with the host materials and create hurdles in fast adsorption/desorption kinetics. With the help of metal-free reagents such as triflic acid, Bronsfed acid, the desorption of chemically adsorbed hydrogen molecules from boron nitride nanotubes was made possible at close to ambient temperature [172]. Recently, Banerjee et al. investigated the poly-lithiated (CLi2) functionalized BNNTs for hydrogen storage application. Functionalization with different concentrations of CLi2 enhances the $E_{ads}$ between 0.25-0.35 eV per H2 with an uptake capacity of 4.41 wt% [173]. The hydrogen storage properties of some of the expanded, doped decorated boron nitride nanostructures are listed above in Table 3.

## CONCLUDING REMARKS

This review article summarizes the progress made in utilizing the carbon and boron nitride nanostructures for hydrogen molecule storage employing a first principle study. As revealed from the theoretical studies, hydrogen molecules are weakly bonded in pristine carbon and boron nitride nanomaterials. Alternative approaches like creating a vacancy, doping, and decoration of atoms by applying strain, an external electric field on the carbon and boron nitride nanomaterials, are promising to enhance hydrogen adsorption ability. Theoretical results suggest that hydrogen molecules are adsorbed on the host materials through physisorption, chemisorption processes. In doping and decoration of transition metals on carbon and boron nitride nanostructures, the d-orbital electrons of TMs hybridize with the sigma-orbital of hydrogen molecules. Such kind of interaction is termed Kubas-interaction. The host materials are decorated with lightweight alkali and alkaline earth metals atoms to overcome the low adsorption energy crisis. Another approach where poly-lithiated molecules are decorated on pristine carbon and boron nitride nanostructures tunes the adsorption/desorption kinetics of the host materials. The alkali and alkaline earth metals, when decorated, transfer charge to the host materials and becomes cationic leading to a higher rate of $H_2$ adsorption. Carbon and oxygen doped on boron nitride nanostructures also show a promising response towards hydrogen storage. Most of the theoretical calculation on the carbon and boron nitride nanostructures shows a potential to be utilized for onboard hydrogen storage for mobile application. However, only a few experimental studies for hydrogen storage have been realized at present. So, it's the right moment to take these theoretical works as guidance to experimentally synthesize the hydrogen storage material for mobile applications at ambient conditions.

## CONFLICT OF INTEREST

None Declare

# ACKNOWLEDGEMENT

D. P. Rai acknowledges core research grant from the Department of Science and Technology SERB (CRG DST-SERB, New Delhi, India) via sanction no. CRG/2018/00009(VER-1).

[133] H. WANG, X. CHENG, H. ZHANG, and Y. TANG, "VERY HIGH HYDROGEN STORAGE CAPACITY OF AL-ADSORBED SINGLE-WALLED CARBON NANOTUBE (SWCNT): MULTI-LAYERED STRUCTURE OF HYDROGEN MOLECULES," *Int. J. Mod. Phys. B*, vol. 27, no. 14, p. 1350061, Jun. 2013, doi: 10.1142/S0217979213500616.

[134] V. J. Surya, K. Iyakutti, M. Rajarajeswari, and Y. Kawazoe, "First-Principles Study on Hydrogen Storage in Single Walled Carbon Nanotube Functionalized with Ammonia," *J. Comput. Theor. Nanosci.*, vol. 7, no. 3, pp. 552–557, Mar. 2010, doi: 10.1166/jctn.2010.1393.

[135] M. Topsakal, E. Aktürk, and S. Ciraci, "First-principles study of two- and one-dimensional honeycomb structures of boron nitride," *Phys. Rev. B - Condens. Matter Mater. Phys.*, vol. 79, no. 11, p. 115442, Mar. 2009, doi: 10.1103/PhysRevB.79.115442.

[136] A. I. Siahlo et al., "Structure and energetics of carbon, hexagonal boron nitride, and carbon/hexagonal boron nitride single-layer and bilayer nanoscrolls," *Phys. Rev. Mater.*, vol. 2, no. 3, 2018, doi: 10.1103/PhysRevMaterials.2.036001.

[137] B. Chettri, P. K. Patra, N. N. Hieu, and D. P. Rai, "Hexagonal boron nitride (h-BN) nanosheet as a potential hydrogen adsorption material: A density functional theory (DFT) study," *Surfaces and Interfaces*, vol. 24, p. 101043, Jun. 2021, doi: 10.1016/j.surfin.2021.101043.

[138] Q. Weng et al., "Tuning of the Optical, Electronic, and Magnetic Properties of Boron Nitride Nanosheets with Oxygen Doping and Functionalization," *Adv. Mater.*, vol. 29, no. 28, p. 1700695, Jul. 2017, doi: 10.1002/adma.201700695.

[139] B. Baumeier, P. Krüger, and J. Pollmann, "Structural, elastic, and electronic properties of SiC, BN, and BeO nanotubes," *Phys. Rev. B - Condens. Matter Mater. Phys.*, vol. 76, no. 8, Aug. 2007, doi: 10.1103/PhysRevB.76.085407.

[140] M. Serhan et al., "The electronic properties of different chiralities of defected boron nitride nanotubes: Theoretical study," *Comput. Condens. Matter*, vol. 22, p. e00439, Mar. 2020, doi: 10.1016/j.cocom.2019.e00439.

[141] S. H. Jhi, D. J. Roundy, S. G. Louie, and M. L. Cohen, "Formation and electronic properties of double-walled boron nitride nanotubes," *Solid State Commun.*, vol. 134, no. 6, pp. 397–402, May 2005, doi: 10.1016/j.ssc.2005.02.007.

[142] W. Lei et al., "Oxygen-doped boron nitride nanosheets with excellent performance in hydrogen storage," *Nano Energy*, vol. 6, pp. 219–224, May 2014, doi: 10.1016/j.nanoen.2014.04.004.

[143] J. Zhou, Q. Wang, Q. Sun, P. Jena, and X. S. Chen, "Electric field enhanced hydrogen storage on polarizable materials substrates," *Proc. Natl. Acad. Sci. U. S. A.*, vol. 107, no. 7, pp. 2801–2806, Feb. 2010, doi: 10.1073/pnas.0905571107.

[144] B. Chettri, P. K. Patra, S. Srivastava, Lalhriatzuala, L. Zadeng, and D. P. Rai, "Electronic Properties of Hydrogenated Hexagonal Boron Nitride (h-BN): DFT Study," *Senhri J. Multidiscip. Stud.*, vol. 4, no. 2, pp. 72–79, Dec. 2019, doi: 10.36110/sjms.2019.04.02.008.

[145] C. Tang, X. Zhang, and X. Zhou, "Most effective way to improve the hydrogen storage abilities of Na-decorated BN sheets: Applying external biaxial strain and an electric field," *Phys. Chem. Chem. Phys.*, vol. 19, no. 7, pp. 5570–5578, Feb. 2017, doi: 10.1039/c6cp07433b.

[146] B. Chettri, P. K. Patra, and D. P. Rai, "Enhanced H$_2$ storage capacity of the bilayer hexagonal Boron Nitride(h-BN) incorporating Van der Waals interaction under applied external electric field," May 2021, Accessed: Jun. 10, 2021. [Online]. Available: http://arxiv.org/abs/2105.15070.

[147] P. Banerjee, B. Pathak, R. Ahuja, and G. P. Das, "First principles design of Li functionalized hydrogenated h-BN nanosheet for hydrogen storage," *Int. J. Hydrogen Energy*, vol. 41, no. 32, pp. 14437–14446, Aug. 2016, doi: 10.1016/j.ijhydene.2016.02.113.

[148] Y. Wang, F. Wang, B. Xu, J. Zhang, Q. Sun, and Y. Jia, "Theoretical prediction of hydrogen storage on Li-decorated boron nitride atomic chains," *J. Appl. Phys.*, vol. 113, no. 6, p. 64309, Feb. 2013, doi: 10.1063/1.4790868.
Page | 29